\documentclass[apj]{emulateapj} \usepackage{times} 

\usepackage{amsmath}
\usepackage{natbib}
\def\be{\begin{equation}}
\def\ee{\end{equation}}
\def\ba{\begin{eqnarray}}
\def\ea{\end{eqnarray}}

\def\lsim{\raise0.3ex\hbox{$\;<$\kern-0.75em\raise-1.1ex\hbox{$\sim\;$}}}
\def\gsim{\raise0.3ex\hbox{$\;>$\kern-0.75em\raise-1.1ex\hbox{$\sim\;$}}}

\def\eps{\varepsilon}
\def\theta{\vartheta}

\begin{document}

\title{Nuclear enhancement of the photon yield in cosmic ray interactions}

\author{Michael Kachelrie\ss\altaffilmark{1}, Igor V.~Moskalenko\altaffilmark{2}
and Sergey S.~Ostapchenko\altaffilmark{2,3}}

\affil{$^1$Institutt for fysikk, NTNU, 7491 Trondheim, Norway}
\affil{$^2$Hansen Experimental Physics Laboratory \& Kavli Institute for Particle Astrophysics and Cosmology, Stanford University, Stanford, CA 94305, U.S.A}
\affil{$^3$Skobeltsyn Institute of Nuclear Physics, Moscow State University, 119991 Moscow, Russia}

\begin{abstract} 
The concept of the nuclear enhancement factor has been used since the beginning of  
$\gamma$-ray astronomy. It provides a simple and convenient way to account for the
contribution of nuclei ($A>1$) in cosmic rays (CRs) and in the interstellar medium (ISM) to the diffuse $\gamma$-ray emission.
An accurate treatment of the dominant emission process, such as hadronic interactions of CRs with the ISM, 
enables one to study CR acceleration processes, CR propagation in the ISM, and provides 
a reliable background model for searches of new phenomena. 
The {\it Fermi} Large Area Telescope ({\it Fermi}-LAT) launched in 2008 provides excellent quality data
in a wide energy range 30 MeV -- 1 TeV where the diffuse emission accounts for the majority of photons.
Exploiting its data to the fullest requires a new study of the processes of $\gamma$-ray production in hadronic interactions.
In this paper we point out that several commonly used studies of the nuclear enhancement factor miss to account
for the spectrally averaged energy loss fraction
which ensures 
that the energy fraction 
transferred to photons is averaged properly 
with the spectra of CR species. 
We present a new calculation of the spectrally averaged energy loss fraction
and the nuclear enhancement factor using
the QGSJET-II-04 and EPOS-LHC interaction models.
\end{abstract}


\section{Introduction}

Launched in 2008, the $\gamma$-ray telescope
{\it Fermi}-LAT provides excellent statistics together with 
superior angular and energy resolution in a wide energy range from 30 MeV -- 1 TeV \citep{2009ApJ...697.1071A}. 
This energy range is dominated by the diffuse Galactic emission, which is the brightest source on the $\gamma$-ray sky. 
Studies of the diffuse $\gamma$-ray emission and extended sources provide invaluable information
about CR intensities and spectra in distant locations. Understanding
the diffuse emission enables us to study particle acceleration processes, CR propagation
in the ISM, and disentangle new phenomena and/or exotic signals 
\citep{2007ARNPS..57..285S,2010ApJ...724.1044S,2012ApJ...750....3A}.

The continuous $\gamma$-ray emission is generated mainly through the decay of neutral pions and
kaons produced in hadronic CR interactions with the
ISM, inverse Compton scattering of CR electrons off interstellar photons, and bremsstrahlung.
The nuclear component of CRs is dominated by protons, but heavier nuclei also provide an essential contribution
to the $\gamma$-ray yield. The latter depends
on the energy range and on the spectra of the CR species.
However, CR spectra and abundances could vary in different locations making an accurate 
evaluation of their contribution to the $\gamma$-ray yield rather difficult.

In all studies of the diffuse $\gamma$-ray emission, the effects of heavier nuclei ($A>1$)
in CRs and in the target material are usually taken into account by simply rescaling
the $\gamma$-ray yield from $pp$-interactions to the CR-ISM $\gamma$-ray yield with a so-called
nuclear enhancement factor $\eps_{\rm M}$.
While such a rescaling is a convenient approximation, application of a single enhancement factor
in many cases could result in significant errors. In fact, there is no a universal enhancement factor
as the rescaling factor depends on the abundances of CRs and the ISM, on the individual spectral shapes of CR species,
as well as on the kinematics of the processes involved, e.g., $pA$ vs.\ $Ap$ yields.

$\gamma$-ray production in $pp$-interactions has been studied in the past using model fits to the data 
\citep{1973ApJ...185..499S,1989cgrc.conf...85S,1981Ap&SS..76..213S,1986ApJ...307...47D,1986A&A...157..223D},
and Monte Carlo simulations \citep{1997ApJ...478..225M,2009APh....31..341M,2006ApJ...647..692K,2012PhRvD..86d3004K}.
The values of the nuclear enhancement factor derived by different authors vary from 1.45 -- 2.0, due to the 
differences in the description of $pp$-interactions, nuclei abundances, and the scaling formalism.
The dependence of $\eps_{\rm M}$ on the 
spectral shapes of CR species was always neglected,
except for a trivial dependence on the relative abundances of CR nuclei.
Since the $\gamma$-ray data become rather accurate, a new study of the nuclear enhancement factor is warranted.

In this work we study how the spectral shape of the CR species and the kinematics of the processes affect
$\eps_{\rm M}$. We use the QGSJET-II-04 event generator, which 
accurately reproduces accelerator data \citep{2011PhRvD..83a4018O}, to simulate $pp$-, $pA$-, and $AA$-interactions, and
compare the results with the most recent calculation by \citet{2009APh....31..341M}
and with another event generator EPOS-LHC \citep{2013arXiv1306.0121P} tuned to LHC data.

\section{Nuclear enhancement factor}

The photon yield $q^{ij}_\gamma(E_\gamma)$ from scattering of CR species of type $i$ 
with differential intensity\footnote{Throughout the paper, $E$ denotes the energy per nucleon.
}
$I_i(E)$ on a target of type $j$ of density $n_j$ is given by 
\be
\label{qij}
 q^{ij}_\gamma(E_\gamma) = n_j \int_{E_\gamma}^\infty d E \;
 \frac{d\sigma^{ij\rightarrow \gamma}(E,E_\gamma)}
		 {d E_\gamma} \: I_i(E),
\ee
where $d\sigma^{ij\rightarrow \gamma}(E,E_\gamma)/d E_\gamma$ is the differential 
inclusive cross section for photon
production. For a power-law spectrum, $I_i(E)=K_i\, E^{-\alpha_i}$,
introducing the energy fraction taken by gammas, $z=E_\gamma/E$,
and the spectrally averaged moment
\be 
\label{Z_spec}
 Z^{ij}_\gamma(E_\gamma,\alpha) = \int_0^1 d z \, z^{\alpha-1}\,
 \frac{d\sigma^{ij\rightarrow \gamma}(E_\gamma/z,z)}{d z}, 
\ee
we can rewrite the photon yield from channel $ij$ as
\be 
\label{q_ij-Z}
 q^{ij}_\gamma(E_\gamma) = n_j \, I_i(E_\gamma) \, Z^{ij}_\gamma(E_\gamma,\alpha _i) .
\ee
Note that in Eq.~(\ref{q_ij-Z}) 
we evaluate the CR intensity $I_i(E)$ at the photon energy $E_\gamma$.

To compare with the most recent approach of \citet{2009APh....31..341M}, 
we can factorize out 
the inelastic cross section $\sigma^{ij}_{\rm inel}(E)$ and the 
photon multiplicity $N^{ij}_{\gamma}(E)$ from the definition of the moment,
i.e., we define\footnote{Where we also formally use $E=E_\gamma$.}
\be 
\label{Z_fact}
\tilde{Z}^{ij}_\gamma(E_\gamma,\alpha) =
\frac{Z^{ij}_\gamma(E_\gamma,\alpha)}{\sigma^{ij}_{\rm inel}(E_\gamma)\,N^{ij}_{\gamma}(E_\gamma)}, 
\ee
with 
\be 
\label{N_gamma}
N^{ij}_{\gamma}(E) = \int_0^1 d z\;f_{ij\rightarrow \gamma}(E,z) .
\ee
Here we introduced also the normalized (per inelastic event) photon energy distribution
\be 
\label{f_gamma}
f_{ij\rightarrow \gamma}(E,z) =
\frac{1}{\sigma^{ij}_{\rm inel}(E)} \,
\frac{d\sigma^{ij\rightarrow \gamma}(E,z)}{d z} . 
\ee
If the inclusive photon cross section satisfied Feynman scaling,
\be 
\label{scaling}
\frac{d\sigma^{ij\rightarrow \gamma}(E,z)}{d z}=F(z),
\ee
$\tilde{Z}^{ij}=1$ would hold for the particular case $\alpha=1$;
on the other hand, for $\alpha=2$, $\tilde{Z}^{ij}$ would correspond to the 
average energy fraction taken by a produced photon (c.f.\ Eqs. [\ref{Z_spec}],
[\ref{Z_fact}-\ref{scaling}]).

We can now rewrite the photon yield from a channel $ij$ as
\be 
\label{q-fact}
 q^{ij}_\gamma(E_\gamma) = n_j \, I_i(E_\gamma) \, \sigma^{ij}_{\rm inel}(E_\gamma) \, 
 N_\gamma^{ij}(E_\gamma) \, \tilde{Z}^{ij}_\gamma(E_\gamma,\alpha _i) \,.
\ee
It is easy to see from Eq.\ (\ref{q-fact}) that the photon yield is not
just proportional to the inelastic cross section $\sigma^{ij}_{\rm inel}(E)$ 
and the number $N_\gamma^{ij}(E)$ of photons produced per interaction, but
depends rather on the spectrally averaged energy fraction 
transferred to photons -- via the ``$Z$-factors''
defined in Eqs.~(\ref{Z_spec}) and (\ref{Z_fact}).
Thus, the yield generally depends on both, the production spectrum of
photons from a channel $ij$ and
the spectrum of CR species $I_i(E)\propto E^{-\alpha _i}$, -- the steeper is the
spectrum and the smaller is the 
average energy fraction $\langle z\rangle$ transferred to photons, 
the smaller is $\tilde{Z}^{ij}_\gamma(E_\gamma,\alpha _i)$ 
and thus the photon yield.

The nuclear enhancement factor $\eps_{\rm M}$ 
due to the admixture of nuclei in CRs and in the ISM is determined by
\ba
\label{eps}
\eps_{\rm M} & =& 1 + \sum_{i+j>2} \eps_{ij} 
=1 + \sum_{i+j>2} \frac{n_j \, I_i(E_\gamma) \, Z^{ij}_\gamma(E_\gamma,\alpha _i)}
{n_p \, I_p(E_\gamma) \, Z^{pp}_\gamma(E_\gamma,\alpha _p)} \nonumber \\
& = & 1 + \sum_{i+j>2} \frac{n_j \, I_i(E_\gamma)}{n_p \, I_p(E_\gamma)} \, 
 \frac{\sigma^{ij}_{\rm inel}}{\sigma^{pp}_{\rm inel}} 
 \frac{N_\gamma^{ij}}{N_\gamma^{pp}} \, 
 \frac{\tilde{Z}^{ij}_\gamma(E_\gamma,\alpha_i)}
	 {\tilde{Z}^{pp}_\gamma(E_\gamma,\alpha_p)} \\
	 &=& 1 + \sum_{i+j>2} \frac{n_j \, I_i(E_\gamma)}{n_p \, I_p(E_\gamma)} \: 
	 m^{\gamma}_{ij}(E_\gamma)\: C_{ij}(E_\gamma,\alpha_i,\alpha_p), \nonumber
\ea
where we introduced also the individual contributions
$\eps_{ij}(E_\gamma) = q^{ij}_\gamma(E_\gamma)/q^{pp}_\gamma(E_\gamma)$ of each channel
to $\eps_{\rm M}$, the ratio of inelastic cross 
sections and multiplicities
\be 
\label{m_ij}
 m^{\gamma}_{ij}(E) = \frac{\sigma^{ij}_{\rm inel}(E)}{\sigma^{pp}_{\rm inel}(E)} 
 \frac{N_\gamma^{ij}(E)}{N_\gamma^{pp}(E)} ,
\ee
and the ratio of the $Z$-factors 
$C_{ij}(E_\gamma,\alpha_i,\alpha_p)=\tilde{Z}^{ij}_\gamma(E_\gamma,\alpha_i)/
\tilde{Z}^{pp}_\gamma(E_\gamma,\alpha_p)$.
Note that the correction 
factors $C_{ij}$
which depend both on the energy distribution of the produced photons
and on the slopes of the primary CR spectra
were missing in the definition
of $\eps_{\rm M}$ used by \citet{2009APh....31..341M}.
As a consequence, the contributions of CR nuclei with $A>1$ 
to the nuclear enhancement factor should deviate from the results
obtained in that study. 
Indeed, as noticed above, the correction factors $C_{ij}$ disappear from
Eq.\ (\ref{eps}) only for the (unrealistic) assumption of the validity of Feynman scaling
and for the (impractical) case of $\alpha =1$. On the other hand,
for steeply falling spectra, such as in the case of Galactic CRs, 
$\alpha\gg 1$, the region of large $z$ gives the dominant contribution to the 
integral defining $Z^{ij}_\gamma(E_\gamma,\alpha)$, i.e.\ it is the photon spectral
shape in the very forward direction,
rather than the photon multiplicity $N_\gamma^{ij}$, which dominates $Z^{ij}_{\gamma}$. 

To illustrate the latter point, let us compare 
 the factors $m^{\gamma}_{ij}(E)$ (Eq.\ [\ref{m_ij}])
and the ratios $Z^{ij}_{\gamma}(E_\gamma,\alpha)/Z^{pp}_{\gamma}(E_\gamma,\alpha)$ 
for $\alpha\gg 1$, for the cases of nucleus-proton ($j=p$) and proton-nucleus
($i=p$) interactions. 
While $m^{\gamma}_{pj}=m^{\gamma}_{jp}$ by virtue of the Lorentz invariance,
the behavior of $Z^{ip}_{\gamma}$ can be understood from the well-known
relation \citep[see][]{Bialas1976}
for the mean number of interacting (``wounded'') projectile nucleons
$\langle n^{ij}_{{\rm w}_{\rm p}}\rangle$ in nucleus-nucleus collisions
\be
\langle n^{ij}_{{\rm w}_{\rm p}}(E)\rangle
=\frac{i\:\sigma^{pj}_{\rm inel}(E)}{\sigma^{ij}_{\rm 
inel}(E)},
\ee
which holds both in the Glauber approach 
and in the Reggeon Field Theory, if one neglects the contribution of target
diffraction, as demonstrated by \citet{1993PAN....56..346K}. 
This leads, in turn, to an approximate
superposition picture for the forward ($z\rightarrow 1$) spectra of secondary photons,
\ba
\label{superp}
\frac{d\sigma^{ij\rightarrow \gamma}(E,z)}{d z}
& =&\sigma^{ij}_{\rm inel}(E) f_{ij\rightarrow \gamma}(E,z)\nonumber\\
&\underset{z\rightarrow1}{\rightarrow}& \sigma^{ij}_{\rm inel}(E) 
\left[\langle n^{ij}_{{\rm w}_{\rm p}}(E)\rangle f_{pj\rightarrow \gamma}(E,z)\right] \\
&=& i\, \frac{d\sigma^{pj\rightarrow \gamma}(E,z)}{d z} , \nonumber
\ea
which thus gives $Z^{jp}_{\gamma}/Z^{pp}_{\gamma}\simeq j >m^{\gamma}_{pj}$ 
for $\alpha _j= \alpha_p=\alpha \gg 1$ (c.f.\ Eq.\ [\ref{Z_spec}]).
 On the other hand, 
assuming that in proton-nucleus and proton-proton interactions 
the shapes of the photon production spectra are similar
in the forward direction, i.e.\ 
$f_{pj\rightarrow \gamma}(E,z)\simeq f_{pp\rightarrow \gamma}(E,z)$
at large $z$, one obtains\footnote{In reality, 
$f_{pj\rightarrow \gamma}(E,z)$ becomes smaller than 
$f_{pp\rightarrow \gamma}(E,z)$ at $z\rightarrow 1$, which may lead 
to a further decrease for
the ratio $Z^{pj}_{\gamma}/Z^{pp}_{\gamma}$ 
in the large $\alpha$ limit, compared to Eq.\ (\ref{Z-pA}), 
though precise results are model-dependent \citep[see the
discussion by][]{2012PhRvD..86d3004K}.} 
\be 
\label{Z-pA}
\frac{Z^{pj}_{\gamma}}{Z^{pp}_{\gamma}}\simeq 
\frac{\sigma^{pj}_{\rm inel}}{\sigma^{pp}_{\rm inel}}<m^{\gamma}_{pj}\, .
\ee
Thus, CR nuclei generally provide a larger contribution to the 
nuclear enhancement factor $\eps_{\rm M}$, compared to previous calculations
based on $m^{\gamma}_{ij}$, while the opposite is true for the contribution
of nuclear species from the ISM.

\section{Numerical results}

The normalized $Z$-factors $\tilde{Z}^{ij}_\gamma(E_\gamma,\alpha)$ were calculated using the QGSJET-II-04 model by \citet{2011PhRvD..83a4018O}.
Table~\ref{tab: z(alpha)} compares the dependence of 
$\tilde{Z}^{ij}_\gamma(E_\gamma,\alpha)$ on the CR spectral index $\alpha$  
for different production channels $ij\rightarrow \gamma$
for two photon energies $E_\gamma=10$ and 100 GeV. Note that 
$\tilde{Z}^{ij}_\gamma$ (c.f.\ Eq.\ [\ref{Z_fact}]) specifies the difference between 
the factor $Z^{ij}_\gamma(E_\gamma,\alpha)$, which defines the partial photon yield
from the channel $ij\rightarrow \gamma$, 
and the product 
$\sigma^{ij}_{\rm inel}(E_\gamma) \, N_\gamma^{ij}(E_\gamma)$. 

It is clear that $\tilde{Z}^{ij}_\gamma$ decreases strongly
for steeper spectral slopes. This is not surprising since
the ratio $Z^{ij}_\gamma(E_\gamma,\alpha)/\sigma^{ij}_{\rm inel}(E_\gamma)$ 
corresponds to a spectrally averaged fraction of the primary energy, $z=E_\gamma/E$,
taken by the produced photons, rather than to the
photon multiplicity -- the steeper is the spectral slope the smaller part
of the very forward production spectrum
of photons $f_{ij\rightarrow \gamma}(z)$ contributes to the integral in 
Eq.\ (\ref{Z_spec}). This explains also why $\tilde{Z}^{ij}_\gamma$ decreases
with energy, especially for large $\alpha$. For relatively small $\alpha$,
the integral in Eq.\ (\ref{Z_spec}) receives a noticeable contribution from
the region of small $z$, which corresponds to the central rapidity plateau 
in the center-of-mass frame for the given process and which is responsible for the
rise of the photon multiplicity $N_\gamma^{ij}(E)$ with energy due to the
violation of Feynman scaling for $f_{ij\rightarrow \gamma}(E,z)$
at small $z$. However, for large $\alpha$ the ratio
$Z^{ij}_\gamma(E_\gamma,\alpha)/\sigma^{ij}_{\rm inel}(E_\gamma)$
is governed by the energy dependence of the production spectrum
$f_{ij\rightarrow \gamma}(E,z)$ at $z\rightarrow 1$, which satisfies 
approximately Feynman scaling. For $\alpha\gg 1$ this 
leads to\footnote{To be more precise, Feynman 
scaling for $f_{ij\rightarrow \gamma}(E,z)$ is (slightly) broken also at
$z\rightarrow 1$, with the spectrum becoming somewhat 
softer at higher energies. This leads to an additional energy decrease 
of $\tilde{Z}^{ij}_\gamma$, compared to Eq.\ (\ref{z(E)}).}
\be 
\label{z(E)}
\frac{\tilde{Z}^{ij}_\gamma(E_2,\alpha)}{\tilde{Z}^{ij}_\gamma(E_1,\alpha)}
\propto \frac{N_\gamma^{ij}(E_1)}{N_\gamma^{ij}(E_2)} \, ,
\ee
i.e.\ $\tilde{Z}^{ij}_\gamma(E_\gamma,\alpha)$ decreases with energy inversely 
proportional to the photon multiplicity in the process.

For practical applications, more important are the ratios 
$Z^{ij}_{\gamma}/Z^{pp}_{\gamma}$ that enter the expressions for
the partial contributions $\eps_{ij}$ to the
nuclear enhancement factor in Eq.\ (\ref{eps}).
The respective results for different production
channels and for different spectral indices calculated with
QGSJET-II-04 are compiled in Table~\ref{tab: z-ratio(alpha)} for $E_\gamma=10$ and 100 GeV;
the corresponding ratios $m^\gamma_{ij}$ of inelastic cross 
sections and multiplicities (Eq.\ [\ref{m_ij}]) are also shown for comparison.
These results confirm our qualitative expectations from the previous
Section -- the actual enhancement factor for He+$p$ collisions, compared
to the $pp$ case, is noticeably higher than estimated from
$m^\gamma_{{\rm He}\,p}$, while for $p$+He interactions the opposite is true. 
Obviously, the discussed trends are stronger for steeper
CR spectra (larger $\alpha$) due to the increasing dominance of the
very forward part of the photon production spectrum.
The same qualitative behavior is observed when comparing the ratios
$Z^{ij}_{\gamma}/Z^{pp}_{\gamma}$ and the factors $m^\gamma_{ij}$,
as calculated using the SIBYLL~2.1 \citep{2009PhRvD..80i4003A} and
EPOS-LHC \citep{2013arXiv1306.0121P} models (Table~\ref{tab: z-ratio(alpha)}), though the
numerical results prove to be quite model-dependent\footnote{A detailed
comparison of different model predictions for photon production with
available accelerator data will be presented elsewhere.}.

Table~\ref{tab: z-factor} shows $Z$-factors $Z^{ij}_{\gamma}$ for various
channels of photon production in CR interactions. For these calculations we use
two up-to-date hadronic interaction models, QGSJET-II-04 and EPOS-LHC.
These results can be used for calculations of the nuclear enhancement
factor when the combined spectrum of a group of CR nuclei can be
approximated by a power-law, $I_i(E)=K_i\, E^{-\alpha_i}$.

As an illustration, we perform a calculation of $\eps_{\rm M}$
in the energy range $E_\gamma=10-1000$ GeV, based on Eq.\ (\ref{eps}), using
the high energy limit of the parametrization of the spectra of groups of CR nuclei
by \citet{2004PhRvD..70d3008H}; the respective parameters $K_i$ and $\alpha_i$ are given 
in Table \ref{tab:para} for convenience.
The values of $\eps_{\rm M}$ are given in Table \ref{tab:enh-factor} for the two interaction models.
As we already emphasized above, our results for partial contributions to the nuclear 
enhancement factor from proton-nucleus ($\eps_{pj}$) and nucleus-proton
($\eps_{ip}$) collisions demonstrate important differences from the approach by \citet{2009APh....31..341M}
and manifest a significant model dependence (c.f.~Table~\ref{tab: z-ratio(alpha)}).
However, the respective corrections work in the \emph{opposite
directions}  and partly compensate each other.
As a consequence, our results for $\eps_{\rm M}$ \emph{in this particular case},
for both interaction models considered,  agree within 5\%
with those of \citet{2009APh....31..341M},  who used a different event generator, DPMJET-III.

\begin{figure}[t]
\center{
\includegraphics[width=0.45\textwidth]{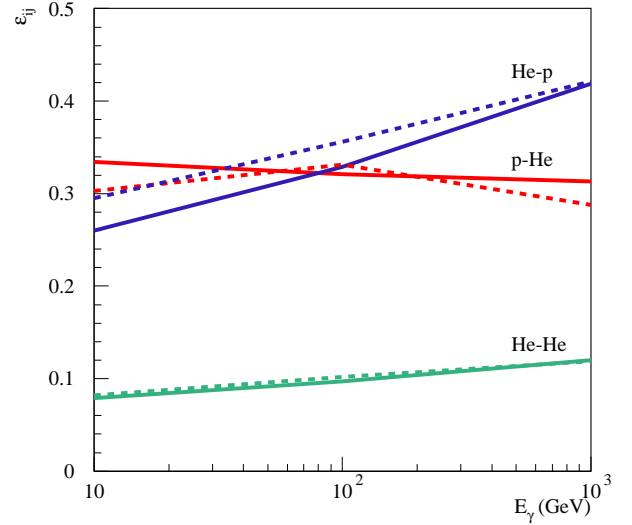}
}
\caption{Partial contributions $\eps_{ij}$ to $\eps_{\rm M}$ for several reaction channels, as indicated in the plot, calculated with QGSJET-II-04 (solid lines)
and EPOS-LHC (dashed lines) models.}
\label{f1}
\end{figure}

Fig.\ \ref{f1} shows the energy dependence of 
the partial contributions $\eps_{ij}$ for $p$+He, He+$p$,
and He+He channels. It is noteworthy that the 
smaller index $\alpha _{\rm He}$ of the He 
component compared to protons has a twofold impact on 
$\eps_{{\rm He}\,p}$ and $\eps_{{\rm He}\,{\rm He}}$:
first, the relative abundance of He increases
with energy, and, second, the respective $Z$-factors become larger for smaller
$\alpha$.

Finally, it is worth stressing that the concept
of the nuclear enhancement factor does not work
in the case of a sharp change in the CR spectral index, as, e.g., around
a spectral break at 230 GV found\footnote{We note that preliminary results from the AMS-02 
experiment (http://www.ams02.org/2013/07/new-results-from-ams-presented-at-icrc-2013/), 
with large statistics, do not show any spectral feature around 230 GV.} 
in the $p$ and He combined data by ATIC-2 
\citep{2009BRASP..73..564P}, CREAM \citep{2011ApJ...728..122Y}, and PAMELA \citep{2011Sci...332...69A}.
In such a case, a direct convolution of the spectra
for different groups of CR nuclei with the respective photon production
distributions, as in Eq.\ (\ref{qij}), is more appropriate. 
Additionally, if such spectral breaks are observed at
different energies per nucleon for different groups of nuclei \citep[e.g.,][]{2011Sci...332...69A}, which is natural to expect from 
rigidity-dependent processes of CRs acceleration and propagation, one may expect
a strong energy dependence of the resulting enhancement factor.

\section{Conclusion}

The concept of the nuclear enhancement factor $\eps_{\rm M}$ provides a simple and convenient way to account for 
the contribution of heavier nuclei in CRs and in the ISM to the diffuse $\gamma$-ray emission. 
The latter is comparable to the contribution of protons, the most abundant species in CRs
and the ISM. 
We have shown that the value of the enhancement depends strongly
on the spectral shapes of CR species: not only via the respective
energy dependence of the partial abundances of primary nuclei,
but also via the spectrally averaged photon energy fraction. It is 
the latter point which was missed in previous calculations.
The provided tables allow a 
calculation of $\eps_{\rm M}$ for an arbitrary composition of CRs and the ISM for a reasonably wide range of power-law indices.
The results for $\eps_{\rm M}$ agree approximately with calculations by \citet{2009APh....31..341M}
for the same spectra of CR species \citep{2004PhRvD..70d3008H}, although we found somewhat larger value of $\eps_{\rm M}$
at energies $E_\gamma>100$ GeV. 

\acknowledgments

IVM and SSO acknowledge support from NASA grants NNX13AC47G and NNX13A092G.

\bibliography{references}

\begin{thebibliography}{22}
\expandafter\ifx\csname natexlab\endcsname\relax\def\natexlab#1{#1}\fi

\bibitem[{{Ackermann} {et~al.}(2012){Ackermann}, {Ajello}, {Atwood}, {Baldini},
  {Ballet}, {Barbiellini}, {Bastieri}, {Bechtol}, {Bellazzini}, {Berenji},
  {Blandford}, {Bloom}, {Bonamente}, {Borgland}, {Brandt}, {Bregeon},
  {Brigida}, {Bruel}, {Buehler}, {Buson}, {Caliandro}, {Cameron}, {Caraveo},
  {Cavazzuti}, {Cecchi}, {Charles}, {Chekhtman}, {Chiang}, {Ciprini}, {Claus},
  {Cohen-Tanugi}, {Conrad}, {Cutini}, {de Angelis}, {de Palma}, {Dermer},
  {Digel}, {Silva}, {Drell}, {Drlica-Wagner}, {Falletti}, {Favuzzi}, {Fegan},
  {Ferrara}, {Focke}, {Fortin}, {Fukazawa}, {Funk}, {Fusco}, {Gaggero},
  {Gargano}, {Germani}, {Giglietto}, {Giordano}, {Giroletti}, {Glanzman},
  {Godfrey}, {Grove}, {Guiriec}, {Gustafsson}, {Hadasch}, {Hanabata},
  {Harding}, {Hayashida}, {Hays}, {Horan}, {Hou}, {Hughes}, {J{\'o}hannesson},
  {Johnson}, {Johnson}, {Kamae}, {Katagiri}, {Kataoka}, {Kn{\"o}dlseder},
  {Kuss}, {Lande}, {Latronico}, {Lee}, {Lemoine-Goumard}, {Longo}, {Loparco},
  {Lott}, {Lovellette}, {Lubrano}, {Mazziotta}, {McEnery}, {Michelson},
  {Mitthumsiri}, {Mizuno}, {Monte}, {Monzani}, {Morselli}, {Moskalenko},
  {Murgia}, {Naumann-Godo}, {Norris}, {Nuss}, {Ohsugi}, {Okumura}, {Omodei},
  {Orlando}, {Ormes}, {Paneque}, {Panetta}, {Parent}, {Pesce-Rollins},
  {Pierbattista}, {Piron}, {Pivato}, {Porter}, {Rain{\`o}}, {Rando}, {Razzano},
  {Razzaque}, {Reimer}, {Reimer}, {Sadrozinski}, {Sgr{\`o}}, {Siskind},
  {Spandre}, {Spinelli}, {Strong}, {Suson}, {Takahashi}, {Tanaka}, {Thayer},
  {Thayer}, {Thompson}, {Tibaldo}, {Tinivella}, {Torres}, {Tosti}, {Troja},
  {Usher}, {Vandenbroucke}, {Vasileiou}, {Vianello}, {Vitale}, {Waite}, {Wang},
  {Winer}, {Wood}, {Wood}, {Yang}, {Ziegler}, \&
  {Zimmer}}]{2012ApJ...750....3A}
{Ackermann}, M., {Ajello}, M., {Atwood}, W.~B., {et~al.} 2012, \apj, 750, 3

\bibitem[{{Adriani} {et~al.}(2011){Adriani}, {Barbarino}, {Bazilevskaya},
  {Bellotti}, {Boezio}, {Bogomolov}, {Bonechi}, {Bongi}, {Bonvicini},
  {Borisov}, {Bottai}, {Bruno}, {Cafagna}, {Campana}, {Carbone}, {Carlson},
  {Casolino}, {Castellini}, {Consiglio}, {De Pascale}, {De Santis}, {De
  Simone}, {Di Felice}, {Galper}, {Gillard}, {Grishantseva}, {Jerse},
  {Karelin}, {Koldashov}, {Krutkov}, {Kvashnin}, {Leonov}, {Malakhov},
  {Malvezzi}, {Marcelli}, {Mayorov}, {Menn}, {Mikhailov}, {Mocchiutti},
  {Monaco}, {Mori}, {Nikonov}, {Osteria}, {Palma}, {Papini}, {Pearce},
  {Picozza}, {Pizzolotto}, {Ricci}, {Ricciarini}, {Rossetto}, {Sarkar},
  {Simon}, {Sparvoli}, {Spillantini}, {Stozhkov}, {Vacchi}, {Vannuccini},
  {Vasilyev}, {Voronov}, {Yurkin}, {Wu}, {Zampa}, {Zampa}, \&
  {Zverev}}]{2011Sci...332...69A}
{Adriani}, O., {Barbarino}, G.~C., {Bazilevskaya}, G.~A., {et~al.} 2011,
  Science, 332, 69

\bibitem[{{Ahn} {et~al.}(2009){Ahn}, {Engel}, {Gaisser}, {Lipari}, \&
  {Stanev}}]{2009PhRvD..80i4003A}
{Ahn}, E.-J., {Engel}, R., {Gaisser}, T.~K., {Lipari}, P., \& {Stanev}, T.
  2009, \prd, 80, 094003

\bibitem[{{Atwood} {et~al.}(2009){Atwood}, {Abdo}, {Ackermann}, {Althouse},
  {Anderson}, {Axelsson}, {Baldini}, {Ballet}, {Band}, {Barbiellini}, \&
  et~al.}]{2009ApJ...697.1071A}
{Atwood}, W.~B., {Abdo}, A.~A., {Ackermann}, M., {et~al.} 2009, \apj, 697, 1071

\bibitem[{{Bia{\l}as} {et~al.}(1976){Bia{\l}as}, {Bleszynski}, \&
  {Czyz}}]{Bialas1976}
{Bia{\l}as}, A., {Bleszynski}, M., \& {Czyz}, W. 1976, \npb, 111, 461

\bibitem[{{Dermer}(1986{\natexlab{a}})}]{1986ApJ...307...47D}
{Dermer}, C.~D. 1986{\natexlab{a}}, \apj, 307, 47

\bibitem[{{Dermer}(1986{\natexlab{b}})}]{1986A&A...157..223D}
---. 1986{\natexlab{b}}, \aap, 157, 223

\bibitem[{{Honda} {et~al.}(2004){Honda}, {Kajita}, {Kasahara}, \&
  {Midorikawa}}]{2004PhRvD..70d3008H}
{Honda}, M., {Kajita}, T., {Kasahara}, K., \& {Midorikawa}, S. 2004, \prd, 70,
  043008

\bibitem[{{Kachelrie{\ss}} \& {Ostapchenko}(2012)}]{2012PhRvD..86d3004K}
{Kachelrie{\ss}}, M., \& {Ostapchenko}, S. 2012, \prd, 86, 043004

\bibitem[{{Kalmykov} \& {Ostapchenko}(1993)}]{1993PAN....56..346K}
{Kalmykov}, N.~N., \& {Ostapchenko}, S.~S. 1993, Physics of Atomic Nuclei, 56,
  346

\bibitem[{{Kamae} {et~al.}(2006){Kamae}, {Karlsson}, {Mizuno}, {Abe}, \&
  {Koi}}]{2006ApJ...647..692K}
{Kamae}, T., {Karlsson}, N., {Mizuno}, T., {Abe}, T., \& {Koi}, T. 2006, \apj,
  647, 692

\bibitem[{{Mori}(1997)}]{1997ApJ...478..225M}
{Mori}, M. 1997, \apj, 478, 225

\bibitem[{{Mori}(2009)}]{2009APh....31..341M}
---. 2009, Astroparticle Physics, 31, 341

\bibitem[{{Ostapchenko}(2011)}]{2011PhRvD..83a4018O}
{Ostapchenko}, S. 2011, \prd, 83, 014018

\bibitem[{{Panov} {et~al.}(2009){Panov}, {Adams}, {Ahn}, {Bashinzhagyan},
  {Watts}, {Wefel}, {Wu}, {Ganel}, {Guzik}, {Zatsepin}, {Isbert}, {Kim},
  {Christl}, {Kouznetsov}, {Panasyuk}, {Seo}, {Sokolskaya}, {Chang}, {Schmidt},
  \& {Fazely}}]{2009BRASP..73..564P}
{Panov}, A.~D., {Adams}, J.~H., {Ahn}, H.~S., {et~al.} 2009, Bulletin of the
  Russian Academy of Science, Phys., 73, 564

\bibitem[{{Pierog} {et~al.}(2013){Pierog}, {Karpenko}, {Katzy}, {Yatsenko}, \&
  {Werner}}]{2013arXiv1306.0121P}
{Pierog}, T., {Karpenko}, I., {Katzy}, J.~M., {Yatsenko}, E., \& {Werner}, K.
  2013, ArXiv: 1306.0121

\bibitem[{{Stecker}(1973)}]{1973ApJ...185..499S}
{Stecker}, F.~W. 1973, \apj, 185, 499

\bibitem[{{Stecker}(1989)}]{1989cgrc.conf...85S}
{Stecker}, F.~W. 1989, in Cosmic Gamma Rays and Cosmic Neutrinos, ed. M.~M.
  {Shapiro} \& J.~P. {Wefel}, 85--119

\bibitem[{{Stephens} \& {Badhwar}(1981)}]{1981Ap&SS..76..213S}
{Stephens}, S.~A., \& {Badhwar}, G.~D. 1981, \apss, 76, 213

\bibitem[{{Strong} {et~al.}(2007){Strong}, {Moskalenko}, \&
  {Ptuskin}}]{2007ARNPS..57..285S}
{Strong}, A.~W., {Moskalenko}, I.~V., \& {Ptuskin}, V.~S. 2007, Annual Review
  of Nuclear and Particle Science, 57, 285

\bibitem[{{Su} {et~al.}(2010){Su}, {Slatyer}, \&
  {Finkbeiner}}]{2010ApJ...724.1044S}
{Su}, M., {Slatyer}, T.~R., \& {Finkbeiner}, D.~P. 2010, \apj, 724, 1044

\bibitem[{{Yoon} {et~al.}(2011){Yoon}, {Ahn}, {Allison}, {Bagliesi}, {Beatty},
  {Bigongiari}, {Boyle}, {Childers}, {Conklin}, {Coutu}, {DuVernois}, {Ganel},
  {Han}, {Jeon}, {Kim}, {Lee}, {Lutz}, {Maestro}, {Malinine}, {Marrocchesi},
  {Minnick}, {Mognet}, {Nam}, {Nutter}, {Park}, {Park}, {Seo}, {Sina},
  {Swordy}, {Wakely}, {Wu}, {Yang}, {Zei}, \& {Zinn}}]{2011ApJ...728..122Y}
{Yoon}, Y.~S., {Ahn}, H.~S., {Allison}, P.~S., {et~al.} 2011, \apj, 728, 122

\end{thebibliography}

\newpage


\begin{deluxetable*}{ccccccc}
\tablecolumns{7}
\tablewidth{0pc}
\tablecaption{
Normalized $Z$-factors $\tilde{Z}^{ij}_\gamma(E_\gamma,\alpha)$ calculated with QGSJET-II-04
\label{tab: z(alpha)}
}
\tablehead{
\colhead{Reaction} & 
\colhead{$\alpha=1.5$} & 
\colhead{$\alpha=2$} & 
\colhead{$\alpha=2.5$} & 
\colhead{$\alpha=3$} &
\colhead{$\alpha=3.5$} & 
\colhead{$\alpha=4$}
}
\startdata
\multicolumn{2}{c}{$E_\gamma=10$ GeV}\\

$p\,p\rightarrow \gamma$ &
  $6.3\cdot 10^{-1}$ & $8.6\cdot 10^{-2}$ & $2.3\cdot 10^{-2}$  &
  $8.3\cdot 10^{-3}$ &  $3.6\cdot 10^{-3}$ &  $1.8\cdot 10^{-3}$ \\
 
$p\,{\rm He}\rightarrow \gamma$ & 
 $6.3\cdot 10^{-1}$ & $8.3\cdot 10^{-2}$ & $2.1\cdot 10^{-2}$  
 & $7.5\cdot 10^{-3}$ &  $3.2\cdot 10^{-3}$  &  $1.6\cdot 10^{-3}$ \\
 
${\rm He}\,p\rightarrow \gamma$ & 
 $6.7\cdot 10^{-1}$ & $9.4\cdot 10^{-2}$ & $2.5\cdot 10^{-2}$  
 & $9.3\cdot 10^{-3}$ &  $4.1\cdot 10^{-3}$  &  $2.1\cdot 10^{-3}$ \\
 
${\rm He}\,{\rm He}\rightarrow \gamma$ & 
 $6.8\cdot 10^{-1}$ & $9.0\cdot 10^{-2}$ & $2.3\cdot 10^{-2}$  
 & $8.4\cdot 10^{-3}$ &  $3.6\cdot 10^{-3}$  &  $1.8\cdot 10^{-3}$ \medskip\\
 
\multicolumn{2}{c}{$E_\gamma=100$ GeV}\\

$p\,p\rightarrow \gamma$ &
 $2.9\cdot 10^{-1}$ & $3.5\cdot 10^{-2}$ & $8.4\cdot 10^{-3}$  &
  $2.8\cdot 10^{-3}$ &  $1.2\cdot 10^{-3}$ &  $5.7\cdot 10^{-4}$ \\
 
$p\,{\rm He}\rightarrow \gamma$ & 
 $2.8\cdot 10^{-1}$ & $3.2\cdot 10^{-2}$ & $7.4\cdot 10^{-3}$  
 & $2.4\cdot 10^{-3}$ &  $1.0\cdot 10^{-3}$  &  $4.8\cdot 10^{-4}$ \\
 
${\rm He}\,p\rightarrow \gamma$ & 
 $3.0\cdot 10^{-1}$ & $3.7\cdot 10^{-2}$ & $9.0\cdot 10^{-3}$  
 & $3.0\cdot 10^{-3}$ &  $1.3\cdot 10^{-3}$  &  $6.2\cdot 10^{-4}$ \\
 
${\rm He}\,{\rm He}\rightarrow \gamma$ & 
 $2.9\cdot 10^{-1}$ & $3.4\cdot 10^{-2}$ & $7.9\cdot 10^{-3}$  
 & $2.6\cdot 10^{-3}$ &  $1.1\cdot 10^{-3}$  &  $5.1\cdot 10^{-4}$ 
\enddata
\end{deluxetable*}

\begin{deluxetable*}{cccccccc}
\tablecolumns{8}
\tablewidth{0pc}
\tablecaption{Ratios $Z^{ij}_{\gamma}/Z^{pp}_{\gamma}$ 
and $m_{ij}^{\gamma}$ factors
for different production channels $ij\rightarrow \gamma$ 
\label{tab: z-ratio(alpha)}}
\tablehead{
 & 
\multicolumn{6}{c}{$Z^{ij}_{\gamma}/Z^{pp}_{\gamma}$}\\
\colhead{Reaction} & 
\colhead{$\alpha=1.5$} & 
\colhead{$\alpha=2$} & 
\colhead{$\alpha=2.5$} & 
\colhead{$\alpha=3$} &
\colhead{$\alpha=3.5$} & 
\colhead{$\alpha=4$} &
\colhead{$m_{ij}^{\gamma}$}
}
\startdata
\multicolumn{4}{c}{QGSJET-II-04: $E_\gamma=10$ GeV}\\

$p\,{\rm He}\rightarrow \gamma$ & 
  $3.77$ & $3.61$ & $3.47$   & $3.40$ &  $3.36$  &  $3.34$ & $3.74$   \\
 
${\rm He}\,p\rightarrow \gamma$ & 
 $4.01$ & $4.11$ & $4.15$   & $4.18$ &  $4.22$  &  $4.27$ & $3.74$  \\
 
${\rm He}\,{\rm He}\rightarrow \gamma$ & 
 $14.0$ & $13.5$ & $13.2$   & $13.0$ &  $12.8$  &  $12.6$ & $12.9$ \medskip\\

\multicolumn{4}{c}{QGSJET-II-04: $E_\gamma=100$ GeV}\\

$p\,{\rm He}\rightarrow \gamma$ & 
 $3.72$ & $3.49$ & $3.38$   & $3.31$ &  $3.26$  &  $3.24$ & $3.85$ \\
 
${\rm He}\,p\rightarrow \gamma$ & 
 $4.04$ & $4.10$ & $4.13$   & $4.14$ &  $4.15$  &  $4.16$ & $3.85$  \\
 
${\rm He}\,{\rm He}\rightarrow \gamma$ & 
 $13.8$ & $13.2$ & $12.8$   & $12.5$ &  $12.3$  &  $12.2$ & $13.7$  \medskip\\

\multicolumn{4}{c}{SIBYLL~2.1: $E_\gamma=100$~GeV}\\

$p\,{\rm He}\rightarrow \gamma$ & 
 $3.54$ & $3.21$ & $3.03$ & $2.91$ & $2.83$ & $2.78$ & $3.71$ \\
 
${\rm He}\,p\rightarrow \gamma$ & 
 $3.71$ & $3.76$ & $3.77$ & $3.77$ & $3.78$ & $3.79$ & $3.71$ \\
 
${\rm He}\,{\rm He}\rightarrow \gamma$ & 
 $11.7$ & $10.7$ & $10.2$ & $9.63$ & $9.35$ & $9.13$ & $12.4$ \medskip\\

\multicolumn{4}{c}{EPOS-LHC: $E_\gamma=100$~GeV}\\

$p\,{\rm He}\rightarrow \gamma$ & 
 $3.60$ & $3.57$ & $3.45$   & $3.33$ &  $3.24$  &  $3.18$ & $4.10$  \\
 
${\rm He}\,p\rightarrow \gamma$ & 
 $3.94$ & $4.20$ & $4.45$   & $4.72$ &  $4.89$  &  $5.12$ & $4.10$ \\
 
${\rm He}\,{\rm He}\rightarrow \gamma$ & 
 $13.5$ & $13.7$ & $13.5$   & $13.3$ &  $13.2$  &  $13.1$ & $14.6$ 
\enddata
\end{deluxetable*}

\begin{deluxetable*}{lccccccc}
\tablecolumns{8}
\tablewidth{0pc}
\tablecaption{$Z$-factors $Z^{ij}_{\gamma}(E_\gamma,\alpha)$ (mbarn)
for different production channels $ij\rightarrow \gamma$ 
\label{tab: z-factor}}
\tablehead{
\colhead{Projectile nucleus} & 
\colhead{Target nucleus} & 
\colhead{$\alpha=2$} & 
\colhead{$\alpha=2.2$} & 
\colhead{$\alpha=2.4$} &
\colhead{$\alpha=2.6$} & 
\colhead{$\alpha=2.8$} &
\colhead{$\alpha=3$}
}
\startdata
\multicolumn{4}{c}{QGSJET-II-04: $E_\gamma=10$ GeV}\\

$p$ ($A$=1) & $p$ & 
 5.45 & 3.06 & 1.84 & 1.17 & 0.771 & 0.529 \\
 
He ($A$=4) & $p$ &
 22.4 & 12.6 & 7.62 & 4.85 & 3.22 & 2.21 \\
 
CNO ($A$=14) & $p$ &
76.8 & 43.8 & 26.6 & 17.1 & 11.4 & 7.89 \\

Mg-Si ($A$=25) & $p$ &
138 & 78.9 & 48.2 & 31.0 & 20.7 & 14.4 \\

Fe ($A$=56) & $p$ &
298 & 171 & 105 & 67.2 & 45.0 & 31.2 \smallskip\\

$p$ ($A$=1) & He & 
 19.7 & 10.9 & 6.48 & 4.07 & 2.68 & 1.83 \\
 
He ($A$=4) & He &
 73.7 & 41.0 & 24.4 & 15.3 & 10.1 & 6.86 \\
 
CNO ($A$=14) & He &
271 & 152 & 91.2 & 57.7 & 38.1 & 26.1 \\

Mg-Si ($A$=25) & He &
473 & 266 & 160 & 101 & 66.8 & 45.7 \\

Fe ($A$=56) & He &
1010 & 569 & 342 & 216 & 143 & 97.5 \medskip\\

\multicolumn{4}{c}{QGSJET-II-04: $E_\gamma=100$ GeV}\\

$p$ ($A$=1) & $p$ & 
 5.93 & 3.20 & 1.86 & 1.14 & 0.736 & 0.492 \\
 
He ($A$=4) & $p$ &
 24.3 & 13.1 & 7.65 & 4.72 & 3.04 & 2.04 \\
 
CNO ($A$=14) & $p$ &
83.3 & 45.4 & 26.6 & 16.5 & 10.7 & 7.21 \\

Mg-Si ($A$=25) & $p$ &
149 & 81.7 & 48.0 & 29.8 & 19.4 & 13.1 \\

Fe ($A$=56) & $p$ &
330 & 181 & 107 & 66.7 & 43.4 & 29.3 \smallskip\\

$p$ ($A$=1) & He & 
 20.7 & 11.0 & 6.33 & 3.85 & 2.45 & 1.63 \\
 
He ($A$=4) & He &
 78.1 & 41.7 & 23.9 & 14.6 & 9.29 & 6.16 \\
 
CNO ($A$=14) & He &
285 & 153 & 88.6 & 54.3 & 34.9 & 23.3 \\

Mg-Si ($A$=25) & He &
506 & 273 & 159 & 97.7 & 63.0 & 42.2 \\

Fe ($A$=56) & He &
1100 & 596 & 346 & 213 & 137 & 92.1 \medskip\\

\multicolumn{4}{c}{QGSJET-II-04: $E_\gamma=1$ TeV}\\

$p$ ($A$=1) & $p$ & 
 6.85 & 3.61 & 2.05 & 1.24 & 0.786 & 0.519 \\
 
He ($A$=4) & $p$ &
 28.4 & 15.0 & 8.51 & 5.14 & 3.26 & 2.14 \\
 
CNO ($A$=14) & $p$ &
95.6 & 50.6 & 28.9 & 17.6 & 11.2 & 7.39 \\
Mg-Si ($A$=25) & $p$ &
174 & 92.4 & 53.0 & 32.3 & 20.6 & 13.6 \\

Fe ($A$=56) & $p$ &
378 & 202 & 117 & 71.3 & 45.7 & 30.5 \smallskip\\

$p$ ($A$=1) & He & 
 23.7 & 12.2 & 6.83 & 4.07 & 2.56 & 1.67 \\
 
He ($A$=4) & He &
 89.2 & 46.1 & 25.8 & 15.4 & 9.66 & 6.31 \\
 
CNO ($A$=14) & He &
321 & 167 & 93.5 & 55.8 & 35.0 & 22.9 \\

Mg-Si ($A$=25) & He &
567 & 296 & 167 & 100 & 63.3 & 41.6 \\

Fe ($A$=56) & He &
1260 & 660 & 375 & 226 & 143 & 94.6 \medskip\\

\multicolumn{4}{c}{EPOS-LHC: $E_\gamma=10$ GeV}\\

$p$ ($A$=1) & $p$ & 
 5.83   & 3.31  & 2.00    & 1.27   & 0.844   &  0.578   \\
He ($A$=4) & $p$ &
 26.0   & 15.0   & 9.27   & 6.00   &  4.04    & 2.82   \\
CNO ($A$=14) & $p$ &
89.6   & 52.3  & 32.4   & 21.1   &  14.3 & 9.99    \\
Mg-Si ($A$=25) & $p$ &
156   & 91.5   & 57.1   & 37.4  &  25.5 & 18.0    \\
Fe ($A$=56)  & $p$ &
342   & 203  & 128   & 84.6   &  58.2     & 41.4   \smallskip\\
 
$p$ ($A$=1) & He & 
 20.7   & 11.4   & 6.68   & 4.12   &  2.64     & 1.75     \\
He ($A$=4) & He &
 82.5   & 46.3   & 27.7   & 17.5   &  11.5     & 7.79  \\
CNO ($A$=14) & He &
309  & 175   & 106   & 67.7   &  44.9    & 30.8   \\
Mg-Si ($A$=25) & He &
562   & 322   & 196   & 126   &  83.7   & 57.6    \\
Fe ($A$=56)  & He &
1200   & 692  & 424  & 273   &  183   & 128   \medskip\\

\multicolumn{4}{c}{EPOS-LHC: $E_\gamma=100$ GeV}\\

$p$ ($A$=1) & $p$ & 
 6.34   & 3.49  & 2.06    & 1.29   & 0.837   &  0.564    \\
He ($A$=4) & $p$ &
 26.6   & 14.9   & 9.01   & 5.75   &  3.84     & 2.66   \\
CNO ($A$=14) & $p$ &
95.4   & 54.9   & 33.8   & 22.0   &  15.0    & 10.6   \\
Mg-Si ($A$=25) & $p$ &
167  & 96.2   & 59.1   & 38.3   &  25.9    & 18.1    \\
Fe ($A$=56)  & $p$ &
373   & 216  & 134   & 87.9   &  60.0    & 42.4  \smallskip\\

$p$ ($A$=1) & He & 
 22.6   & 12.3   & 7.18   & 4.44   &  2.88    & 1.94  \\
He ($A$=4) & He &
 86.5   & 47.2  & 27.7   & 17.2  &  11.2     & 7.60    \\
CNO ($A$=14) & He &
321  & 177  & 105  & 66.3   &  43.7   & 29.9   \\
Mg-Si ($A$=25) & He &
582  & 324  & 193   & 122   &  80.2   & 54.7   \\
Fe ($A$=56)  & He &
1320   & 744  & 449   & 286 &  190   & 130  \medskip\\

\multicolumn{4}{c}{EPOS-LHC: $E_\gamma=1$ TeV}\\

$p$ ($A$=1) & $p$ & 
 7.61   & 4.15  & 2.45    & 1.54  & 1.01  &  0.693    \\
He ($A$=4) & $p$ &
 31.1   & 17.3   & 10.3  & 6.51   &  4.31 
   & 2.96    \\
CNO ($A$=14) & $p$ &
106  & 60.2  & 36.7   & 23.7   &  16.0    & 11.3   \\
Mg-Si ($A$=25) & $p$ &
192  & 110  & 68.2  & 44.7  &  30.6    & 21.8   \\
Fe ($A$=56)  & $p$ &
433   & 253  & 159   & 105   &  73.6    & 53.1   \smallskip\\

$p$ ($A$=1) & He & 
 25.3   & 13.4   & 7.73   & 4.71   &  3.01   & 2.00   \\
He ($A$=4) & He &
98.3  & 53.1  & 31.0  & 19.2  &  12.5     & 8.44    \\
CNO ($A$=14) & He &
360  & 197  & 116  & 72.7  &  47.5   & 32.3   \\
Mg-Si ($A$=25) & He &
654  & 361  & 214  & 135   &  88.7   & 60.6   \\
Fe ($A$=56)  & He &
1480   & 829  & 498   & 317 &  210   & 145 
\enddata
\end{deluxetable*}

\begin{deluxetable*}{cccccc}
\tablecolumns{6}
\tablewidth{0pc}
\tablecaption{Spectral parameterizations for groups of CR nuclei
\citep{2004PhRvD..70d3008H}
\label{tab:para}}
\tablehead{
& \multicolumn{5}{c}{Groups of nuclei}\\
\colhead{Parameters} & 
\colhead{H ($A$=1)} & 
\colhead{He ($A$=4)} & 
\colhead{CNO ($A$=14)} & 
\colhead{Mg-Si ($A$=25)} &
\colhead{Fe ($A$=56)}
}
\startdata
$K$ & 14900 & 600 & 33.2 & 34.2 & 4.45 \\

$\alpha$ & 2.74 & 2.64 & 2.60 & 2.79 & 2.68 
\enddata
\end{deluxetable*}

\begin{deluxetable}{lccc}
\tablecolumns{4}
\tablewidth{0pc}
\tablecaption{
Nuclear enhancement factors $\eps_{\rm M}$ calculated for CR composition
given in Table~\ref{tab:para} 
\label{tab:enh-factor}}
\tablehead{
& \multicolumn{3}{c}{Photon energy, GeV}\\
\colhead{Models} & 
\colhead{10} & 
\colhead{100} & 
\colhead{1000} 
}
\startdata
QGSJET-II-04 & 1.85   & 1.95  & 2.09 \\
EPOS-LHC & 1.88   &  2.02  &  2.09
\enddata
\end{deluxetable}

\end{document}